\documentstyle[prb,aps,epsf,floats]{revtex}

\begin{document}

\draft

\twocolumn[\hsize\textwidth\columnwidth\hsize\csname@twocolumnfalse%
\endcsname
\title{The critical exponents of the two-dimensional 
Ising spin glass revisited:\\
Exact Ground State Calculations and Monte Carlo Simulations
}

\author{H. Rieger$^a$, L. Santen$^b$, U. Blasum$^c$\\}
\address{$^a$ HLRZ c/o Forschungszentrum J\"ulich, 52425 J\"ulich, Germany}
\address{$^b$ Institut f\"ur Theoretische Physik, Universit\"at zu K\"oln,
50937 K\"oln, Germany}
\address{$^c$ Zentrum f\"ur Paralleles Rechen, Universit\"at zu K\"oln,
50937 K\"oln, Germany}
\author{ M. Diehl$^{d}$, M. J\"unger$^{d}$\\}
\address{$^d$ Institut f\"ur Informatik, Universit\"at zu K\"oln,
50937 K\"oln, Germany}

\date{\today}

\maketitle

\begin{abstract}
  The critical exponents for $T\to0$ of the two-dimensional Ising spin
  glass model with Gaussian couplings are determined with the help of
  exact ground states for system sizes up to $L=50$ and by a Monte
  Carlo study of a pseudo-ferromagnetic order parameter. We obtain:
  for the stiffness exponent $y(=\theta)=-0.281\pm0.002$, for the
  magnetic exponent $\delta=1.48 \pm 0.01$ and for the chaos exponent
  $\zeta=1.05\pm0.05$. From Monte Carlo simulations we get the thermal
  exponent $\nu=3.6\pm0.2$. The scaling prediction $y=-1/\nu$ is
  fulfilled within the error bars, whereas there is a disagreement
  with the relation $y=1-\delta$.

\end{abstract}
\vskip 0.5 cm
]

\section{Introduction}

It is now widely believed that the bond-disordered two-dimensional
Ising spin glass model with short range interactions does not have a
phase transition at any non-vanishing temperature
\cite{rieger,remark}. At zero temperature the spin glass is in its
ground state (i.e.\ the spin configuration with the lowest possible
energy), which might be degenerate or unique depending on the
probability distribution of the spin interactions.  This ground state
is unstable with respect to thermal fluctuations and any non-vanishing
temperature destroys this long range spin glass order. By decreasing
the temperatures on the other hand the spatial correlations
grow resulting in a divergence of the spin glass susceptibility at
zero temperature. This scenario is characterized by a set of critical
exponents that depend on certain features of the bond
distribution. Experiments on Rb$_2$Cu$_{1-x}$Co$_x$F$_4$ clearly
confirmed this picture \cite{dekker} and reported values for the 
critical exponents, which are compatible with those predicted by the
numerical investigations.

The latter has been pursued in four different ways: Monte Carlo
simulations at finite temperatures \cite{bhatt,liang}, high
temperature series expansion \cite{series}, transfer matrix
calculations \cite{transfer1,domainwall,kawashima2} and exact
determination of ground states via combinatorial optimization
\cite{barahona,grotschel,simone,saul} or replica optimization
\cite{kawashima}. A scaling theory by Bray and Moore \cite{bm_sc}
establishes relations between exponents quantifying the stiffness of
the ground state and the critical exponents characterizing the
temperature dependent divergence of various thermodynamic quantities
like correlation length or susceptibility. 

With the most recent numerical studies a controversy arisen on the
critical exponents of the two-dimensional Ising spin glass with
Gaussian couplings: A Monte Carlo study by Liang \cite{liang} (using a
Swendsen-Wang-type cluster algorithm) and a numerical transfer matrix
calculation by Kawashima et al. \cite{kawashima2} yield a value for
the thermal exponent $\nu$ which is significantly different from the
early estimates \cite{bhatt,series,transfer1,domainwall}. Moreover,
Kawashima et al.\ \cite{kawashima} also study the ground state
magnetization of this model in an external field and report a value
for the magnetic field exponent, which is, using a scaling relation
\cite{bm_sc}, incompatible with the stiffness exponent found in
domain wall renormalization group studies.

This observation and the progress in algorithmic developments
motivated us to revisit the critical exponents of the two-dimensional
Ising spin glass model. In this paper we present a synopsis of a
zero-temperature (ground state) {\it and} a finite-temperature (Monte
Carlo) approach to estimate the numerical values for these critical
exponents. For the former we report results obtained from exact ground
states for the largest system sizes possible to date, resulting in the
most reliable estimates for the stiffness exponent $y$ and magnetic
field exponent $\delta$ reported so far. For the Monte-Carlo
simulations we propose a pseudo-ferromagnetic order parameter that is
defined by a projection of spin configurations onto the exactly know
ground state and show that the thermal exponent $\nu$ is identical to
the values that have been obtained by Bhatt and Young \cite{bhatt}
studying the Edwards-Anderson (EA) order parameter.  In the context of
domain growth and non-equilibrium dynamics this concept has already
been introduced and proven to be useful by direct comparison with the
so called replica overlap \cite{kisker}.

The two-dimensional Ising spin glass model with a Gaussian
distribution of couplings, which we consider throughout this paper, is
defined by the Hamiltonian
\begin{equation}
H = \sum_{<ij>}J_{ij}S_iS_j - h\sum_iS_i\;, \qquad S_i = \pm 1
\label{EA_Ham}
\end{equation}
where $\langle ij\rangle$ denotes all nearest neighbor pairs on a
$L\times L$ square lattice with periodic boundary conditions and the
random interaction strengths obey a Gaussian probability distribution
with mean zero and variance one. The parameter $h$ denotes an external
magnetic field strength.

The outline of the paper is as follows: In the next section we present
our results from exact defect energy calculations, which provides us
with an estimate for the stiffness exponent $y$.  In section
\ref{MC_sec} we present a conventional finite size scaling analysis of
Monte Carlo data. In contrast to earlier investigations we used exact
ground state configurations instead of replica systems in order to
establish an order parameter. Section \ref{Mag_sec} focuses on the
exact calculation of ground state magnetizations in an external field
and its finite size scaling properties. Section \ref{Chaos_sec}
presents a study of the sensitivity of the ground state with respect
to slight perturbations of the coupling strength. The last section is
a summary plus discussion.

\section{Defect Energy}

\label{defect_sec}

\subsection{Scaling Theory}

The scaling theory by Bray and Moore \cite{bm_sc} starts
with a coarse grained picture for the spin interactions. It
hypothesizes the following scaling ansatz for an effective coupling
$\tilde{J}(L)$ among (block) spins on length scale $L$ at an
infinitesimal temperature:
\begin{equation}
  \tilde{J}(L) \sim JL^y
  \label{jeff_scale}
\end{equation}
where $J$ denotes the variance of the original bond distribution.  For
positive stiffness exponent $y$ (sometimes $\theta$) the coupling
becomes stronger on larger length scales, which means that it is
harder to flip collectively a connected set of spins of linear
dimension $L$. Thus thermal fluctuations are irrelevant and the spin
glass ordered state persists at low temperature. A negative exponent
$y$, as we expect for $d=2$, on the other hand indicates the
instability of the spin glass ground state. In this case the spin glass
transition takes place only at zero temperature and $y$ is related to
the thermal exponent $\nu$ determining the divergence of the
correlation length $\xi\sim T^{-\nu}$:

The temperature dependence of the correlation length $\xi$ near $T=0$
can be inferred from equating the two energy scales set by the
effective coupling constant and the temperature.  At low temperatures
where (\ref{jeff_scale}) holds one has then
\begin{equation}
  \xi \sim T^{1/y}
  \label{xi_low}
\end{equation}
and therefore
\begin{equation}
  y=-1/\nu\label{ypred}
\end{equation}
In this way the exponent $y$, which we calculate in this section, has
to be compared with $\nu$ determined in finite temperature Monte Carlo
simulations discussed in the next section.

\subsection{Algorithm and Results}

The problem of finding a spin configuration with lowest energy can be
transformed into the problem of finding a maximum weight cut in a
special weighted graph, that represents the interaction structure of
the spin glass system \cite{bieche}.  This is known under the name
Max-Cut problem and is in general a NP-complete problem \cite{karp}.
If the graph is planar, as in the two-dimensional case with {\it free}
or {\it fixed} boundary conditions, the problem is solvable in
polynomial time \cite{bieche}. If one has periodic or anti-periodic
boundary conditions or if an external field (representable as an extra
node to which all other nodes are connected) is present the graph is
not planar any more even in two dimensions. Hence, the situation we
are studying here is indeed a NP-complete problem.

The results of this section and of sections \ref{Mag_sec} and
\ref{Chaos_sec} are based on the application of a so-called branch \&
cut algorithm to the ground state problem \cite{grotschel}.  This
algorithm always finds an exact ground state of the given spin glass
system. For further details about this algorithm and its
implementation see ref.\ \cite{algorithm}.  An important feature of
this approach is, that the returned solutions are proved to be
optimal. Exact ground states of grid sizes up to $100 \times 100$ can
be determined in a moderate amount of computation time. The $100
\times 100$ instances take between $1.5$ and $8$ hours, $4$ hours on
average. Up to grid sizes of $50 \time 50$ each run takes less than 15
minutes.

The NP-completeness is not a serious problem as long as the
system sizes $L$ one studies are not in the region where the
exponential dominates, and for really large $L$, where it does
dominate, the exponent is very small. In the range $L \leq 32$ our
empirical CPU-times can be fitted by a power law ($\tau \propto
L^{3.5}$). For bigger systems there enters an exponential term
$\approx 1.2^{L}$ inside the used size range. Note that these are only
empirical observations but no rigorous bound for the complexity. For
comparison Kawashima and Suzuki \cite{kawashima} reported about a
replica optimization method which approximates ground states
efficiently. They achieved an average CPU-time $\tau(L)$ for systems
of linear size $L$ which can be fitted by a power law like $\tau
\propto L^{5.3}$ inside the same size range ($L \leq 32$). Although
Kawashima and Suzuki only approximate the ground states while we
always find optimal solutions, their CPU-times are similar to ours.
On average, their biggest systems ($32\times 32$) took 260 seconds,
while we needed 160 seconds for $40\times 40$ spin glasses (1500
seconds for $60\times 60$). Kawashima and Suzuki used a VAX 6440, our
computations were carried out on a SPARC 10/612.

\begin{figure}[hbt]
\epsfxsize=\columnwidth\epsfbox{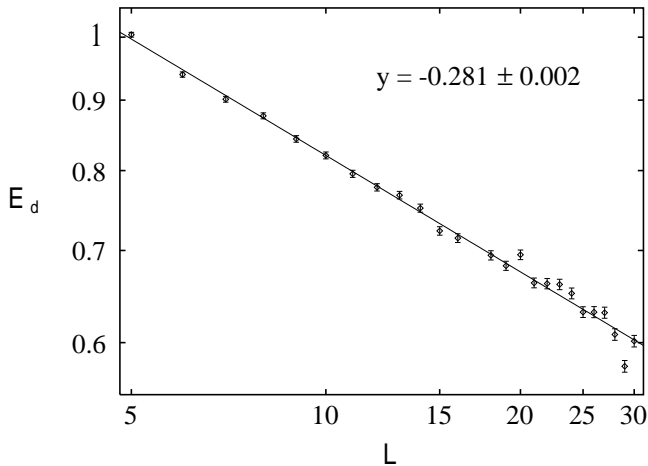}
\caption{Defect energy $E_d$ as a function of the system size $L$ in 
a log-log plot. The straight line is a least square fit giving the
exponent $y=-0.281$.}
\label{defect}
\end{figure}

One can determine the defect energy by investigating the sensitivity
of the ground state energy to boundary conditions\cite{bm_sc,bray88},
which can be quantified by a stiffness exponent measuring the extra
energy of a defect line through the whole sample. The block coupling
$J'$ is then given by $J' = \sqrt{(E_p - E_a)^2}$, where $E_p$ and
$E_a$ are the ground state energies of the system under periodic and
anti-periodic boundary conditions, respectively. We compute this value
in the following way. First we solve the given spin glass system to
optimality under periodic boundary conditions, i.e., we find an exact
ground state configuration $\omega_p$ and its energy $E_p =
E(\omega_p)$. Then we choose two neighboring ``columns'' of spins and
multiply all couplings by $-1$ that link these two spin sets.  By this
modification of the couplings we impose anti-periodic boundary
conditions to the original system. With this slightly changed
objective function we rerun our branch \& cut code to find a ground
state configuration $\omega_a$ with energy $E_a=E(\omega_a)$.

Due to the very small magnitude of $E_a -E_p$ it is necessary to have
a very large number of samples to obtain stable statistics.  For each
size $L \leq 30$ of our $L \times L$ spin glasses we ran 
$\lceil{\frac{2\cdot 10^5}{L}}\rceil$ samples. 
The resulting mean values of the defect
energies versus the system size $L$ are shown in 
\mbox{Fig.\ \ref{defect}}. 
A least square fit yields the value
\begin{equation}
y=-0.281\pm0.002\quad\Rightarrow\quad\nu=3.559\pm0.025
\label{yest}
\end{equation}
The errors are statistical errors only. This estimate agrees roughly
with less accurate early estimates $\nu = 2.96 \pm 0.22$ and $\nu =
4.2 \pm 0.5$ from transfer matrix calculations \cite{transfer1} as
well as $\nu = 3.56 \pm 0.06$ and $\nu = 3.4 \pm 0.1$ from domain wall
renormalization calculations \cite{domainwall}.  Note that Bray and
Moore \cite{bm_sc} report an estimate $y=-0.291 \pm 0.002$, which has
an error bar that is identical to ours. However, their maximum system
size is $L=12$ and they did not calculate {\it exact} ground states.

Our result for $y$ (\ref{yest}) implies a value for $\nu$, if the
scaling prediction (\ref{ypred}) is correct, that differs substantially
from more recent estimates \cite{liang,kawashima2}. Since in these
works the thermal exponent $\nu$ has been determined directly, we
shall do this, too, in the next section.

\section{Monte Carlo results}

\label{MC_sec}

In this section we present our results from finite temperature Monte
Carlo simulations. For this purpose we introduce first a number of
quantities that are of interest for studying the critical properties
of Ising spin glasses in zero external field $h$. 

\subsection{Scaling Relations and Methodology}

A characteristic feature of a spin glass transition at a temperature
$T_c$ (which might be zero) is the divergence of the so called spin glass
susceptibility
\begin{equation}
\chi = \frac{1}{N} \sum_{<ij>}
[\langle S_iS_j\rangle^2]_{\rm av},
\label{sus_def}
\end{equation}
where $[\dots]_{\rm av}$ denotes the average over the quenched
disorder and $\langle\dots\rangle$ a thermal average. Approaching the transition
temperature from the paramagnetic phase one observes
$\chi\sim(T-T_c)^{-\gamma}$, which defines the susceptibility exponent
$\gamma$. As already mentioned there is a diverging length scale at
the transition, the spin glass correlation length
$\xi\sim(T-T_c)^{-\nu}$, which governs the scaling form of the 
correlation function near $T_c$:
\begin{equation}
  G(r)= [<S_iS_{i+r}>^2]_{\rm av} \sim r^{-(d-2+\eta)} \tilde g(r/\xi)
  \label{g_krit}
\end{equation}
Obviously $\gamma = (2-\eta)\nu$. In the case $T_c>0$ there is a
non-vanishing Edwards-Anderson order parameter $q_{EA}=[\langle
S_i\rangle^2]_{\rm av}$ below the transition and one has
$q_{EA} \sim (T_c-T)^{\beta}$ for $T<T_c$, the order parameter
exponent $\beta$ obeying the hyperscaling relation $\beta =
\frac{\nu}{2}(d-2+\eta)$.

In two dimensions $T_c=0$ and since we are concerned with a
continuous bond distribution, in which case the ground state is
non-degenerate, one has
\begin{equation}
 \begin{array}{ccc}
   \eta      &=&0\\
   \beta     &=&0\\
   \gamma/\nu &=&2
 \end{array}
\label{exp0}
\end{equation}
Thus we are left with a single unknown exponent $\nu$, which we
determined from the scaling behavior of the susceptibility and the
Binder cumulant. In contrast to previous investigations we used exact
ground states to define a pseudo-ferromagnetic order parameter via
\begin{equation}
M=[\langle q\rangle ]_{\rm av}
\label{qgrn_def}
\end{equation}
with 
\begin{equation}
q = \frac{1}{N} \sum_{i=1}^N S_i S_i^0\qquad(N=L^2)
\end{equation}
where $S_i^0$ denotes the value of the spin at site $i$ in one of the
two ground state configurations. Note that in contrast to the EA-order
parameter here is only one fluctuating field involved, which would in
principle reduce the order parameter exponent to $\beta/2$. Since we
have $T_c=0$ and thus $\beta=0$ this is not relevant here. The
corresponding order parameter susceptibility is defined via
\begin{equation}
    \chi_L = N [\langle q^2 \rangle]_{\rm av}.
  \label{sus_def2}
\end{equation}
For the finite size scaling form of the susceptibility we expect according 
to (\ref{exp0})
\begin{equation}
\chi_L(T) = L^2\overline{\chi}(L^{1/\nu}T).
\label{chi_fs2d}
\end{equation}

The second quantity we studied was the disorder averaged Binder cumulant
\begin{equation}
g_L = \frac{1}{2}
\left[3-\frac{\langle q^4\rangle}{\langle q^2\rangle^2}
\right]_{\rm av}.
\label{g_def}
\end{equation}
Since this is a dimensionless combination of moments
its finite size scaling form is
\begin{equation}
g_L(T) = \overline{g}(L^{1/\nu}T). 
\label{g_scale}
\end{equation}
This quantity provides us with a second independent estimate for the
scaling exponent $\nu$.

We applied single spin flip Glauber dynamics to perform our
simulations with a spin flip probability given by
\begin{equation}
  w(S_i \rightarrow -S_i) = \frac{1}{1+\exp(\Delta E/T)} ,
\label{flipprob}
\end{equation}
and $\Delta E$ being the energy difference between the new  and the 
old state. Time is measured in Monte Carlo sweeps (MCS) through the whole
 lattice.

\begin{figure}[hbt]
\epsfxsize=\columnwidth\epsfbox{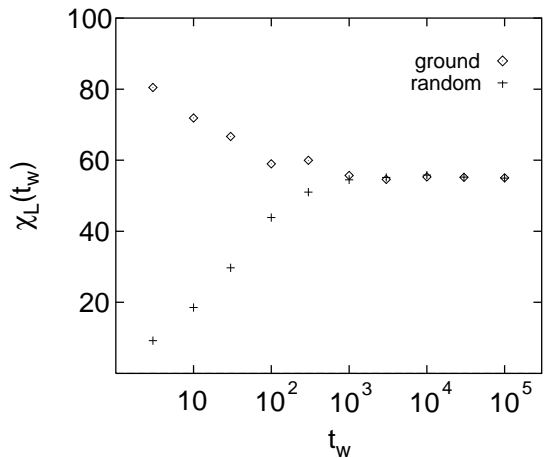}
\caption{Plot of the susceptibility at $T=1.4$ for
$L=10$ averaged over $192$ samples. We compare the
values of $\chi_L(t_w)$ obtained from systems 
initialized with a ground state configuration 
with those obtained from systems with a random initial configuration. 
Obviously the results become time independent 
(aside from statistical fluctuations) if
both estimates agree.}
\label{equi}
\end{figure}

The estimate of the critical exponents necessitates the
determination of the equilibrium values of the thermodynamic
quantities of interest. Due to the slow relaxation of spin glasses
it is difficult to decide whether the values are stationary or not,
because it is hard to discriminate between real- and quasi-stationary
values of the functions. This is why we used a definite criterion
analogous to the criterion introduced by Bhatt and Young \cite{bhatt}:

We simulated two replicas of the system, one which has been
initialized with a random configuration and the other with the
ground state configuration. From Fig.\ref{equi} it can be clearly seen
that if both estimates agree we obtained a time independent value of
the susceptibility, which we took as our equilibration criterion.

\subsection{Results}

We studied the temperature dependent scaling behavior system
sizes extending from $L=6$ to $L=12$. The number of samples is chosen
that approximately $N\cdot \#\mbox{samples} = \mbox{const.}$ holds. We simulated at least
$128$ samples for $L=12$.

\begin{figure}[hbt]
\epsfxsize=\columnwidth\epsfbox{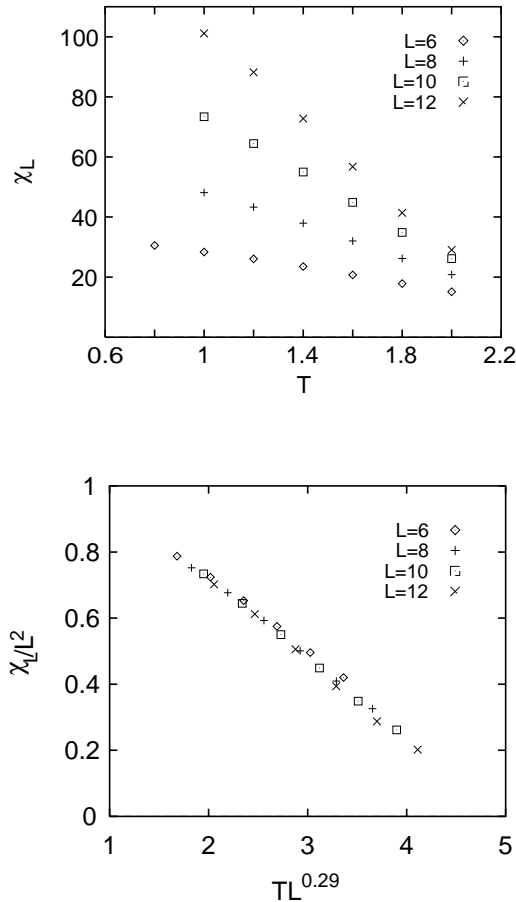}
\caption{Equilibrium values of the susceptibility depending on
temperature and system size.}
\label{sus_res}
\end{figure}
\begin{figure}[hbt]
\epsfxsize=\columnwidth\epsfbox{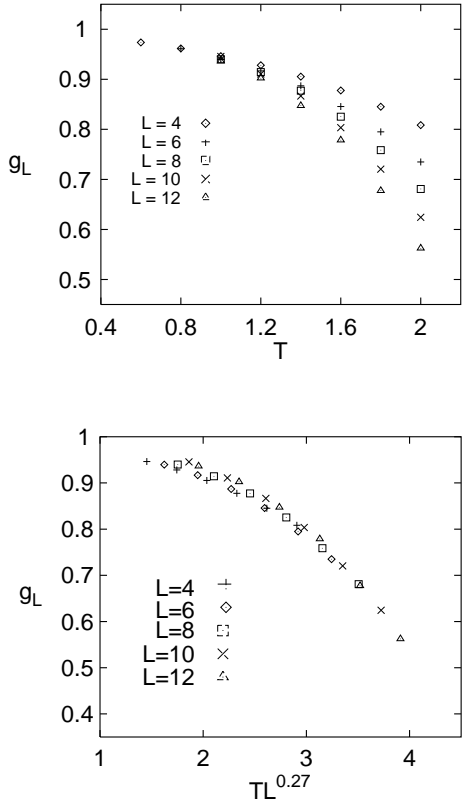}
\caption{Results of $g_L$. Only equilibrium-values are shown. }
\label{g_res}
\end{figure}

Fig.\ref{sus_res} shows the equilibrium values of the susceptibility 
for various system sizes. We could reach the equilibrium 
value of the susceptibility in the chosen time interval 
for $T\geq 1.0$ $(L>6)$ and $T\geq0.8$ $(L=6)$ respectively. 
With this data we got an estimate from the scaling-ansatz 
(\ref{chi_fs2d}). The best data collapse we obtained
(see Fig.\ref{sus_res}) for
\begin{equation}
  \nu =  3.4 \pm 0.2.
  \label{nu_sus}
\end{equation}
The error bars denote the interval of exponents where we get an
indistinguishable data collapse.  

Fig.\ref{g_res} shows the equilibrium values of $g_L$ for the 
same samples. The equilibrium value could be reached within
the same time interval for some smaller temperature. This is why
the data is more sensitive to changes of the value of the 
critical exponent. 
Thus we obtained the best data collapse for
\begin{equation}
  \nu = 3.7 \pm 0.1.
  \label{nu_g}
\end{equation}
in agreement within the error bars with the value determined above.
Concluding we obtain an average value of 
\begin{equation}
  \overline \nu =  3.6 \pm 0.2
  \label{nu_mc_av}
\end{equation}
for the critical exponent from our Monte Carlo simulations. This value
agrees well with the estimate (\ref{yest}) that we obtained from the
defect energy calculations in the last section.  It differs
substantially from the more recent estimates obtained by a cluster
Monte Carlo study \cite{liang} ($\nu=2.0\pm0.2$) and a numerical
transfer matrix calculation \cite{kawashima2} ($\nu=2.08\pm0.01$).

\section{Ground state Magnetization}

\label{Mag_sec}

A nonzero external field $h$ induces a non-vanishing magnetization
$m=N^{-1}\sum_{i=1}^N S_i^0$ in a system with ground state
$\{S_i^0\}$.  The relation between magnetization and field strength is
highly nontrivial in general and motivates the introduction of a new
exponent $\delta$ characterizing this relation in the infinite system
($L\to\infty$) for small fields ($h\ll J$):
\begin{equation}
m_\infty(h)\sim h^{1/\delta}
\label{maginf}
\end{equation}
The corresponding finite size scaling form and a scaling relation
between $\delta$ and the already known exponent $y$ can be obtained by
the following argument \cite{bm_sc}:

If the ground state is non-degenerate the spins are randomly oriented
within an infinitesimal field at $T=0$. Hence the magnetization $m_L$
of a finite system in zero field is a random variable with variance
$1/N$, implying $m_L(h=0)\sim L^{-d/2}$. As a further consequence of
the random orientations the total magnetic moment of a block spin of
linear dimension $L$ is of order $L^{d/2}$, thus the magnetic
field on this length scale has to be rescaled according to
\begin{equation}
  \tilde h(L) \sim L^{d/2} h\;.
  \label{h_scale}
\end{equation}
in contrast to a ferromagnet, where we would have $\tilde{h}(L)\sim L^d h$.
For nonzero field (at $T=0$) one would expect $m_L(h)\cdot L^{d/2}$ to
be a function of the dimensionless ratio of energy scales
$\tilde{J}(L)$ and $\tilde{h}(L)$ only, thus
\begin{equation}
  m_L(h)=L^{-d/2}\,\tilde{m}(L^{d/2-y}\,hJ^{-1})
  \label{m_scale1}
\end{equation}
with $\tilde{m}(x\to0)={\rm const.}$. Since for $L\to\infty$ the
$L$-dependence of the magnetization has to drop out it is
$\tilde{m}(x\to\infty)\sim x^{d/(d-2y)}$.  Moreover, in this limit we have
to recover (\ref{maginf}), which implies for $d=2$
\begin{equation}
\delta=1-y
\label{dypred}
\end{equation}
Rewriting (\ref{m_scale1}) slightly
for our our purposes yields
\begin{equation}
  m_L(h)=L^{-1}\,\overline{m}(Lh^{1/\delta})
  \label{m_scale}
\end{equation}
with $\overline{m}(x\to0)={\rm const.}$ and
$\overline{m}(x\to\infty)\propto x$. Note that (\ref{m_scale}) should
hold independently of the correctness of the above derivation of the
scaling relation (\ref{dypred}): The length scale induced by the
magnetic field is given by $h^{-1/\delta}$ and (\ref{m_scale}) is
simply the finite size scaling form one would expect for the
magnetization.

\begin{figure}[hbt]
\epsfxsize=\columnwidth\epsfbox{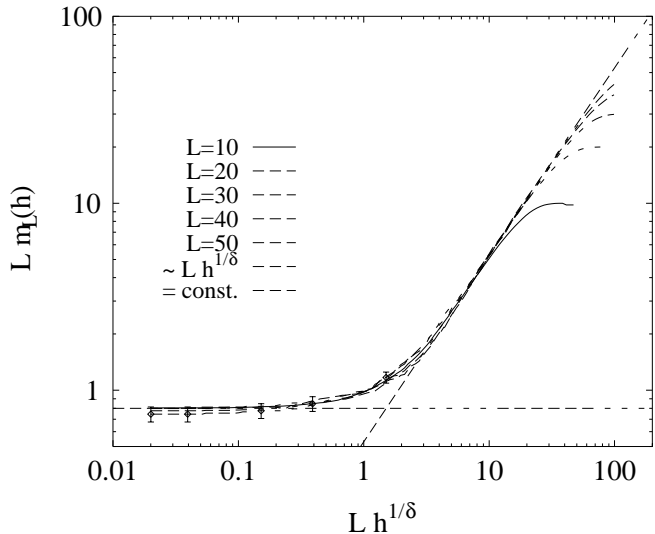}
\caption{Scaling plot for the ground state magnetization:
$Lm_L(h)$ versus $Lh^{1/\delta}$ for various system sizes with
$1/\delta=0.675$. Note that for high fields $h\to\infty$ the curves
have to saturate at $L\cdot m_L(h\to\infty)=L$.}
\label{gr_mag}
\end{figure}

With our branch \& cut algorithm we are not only able to compute
$m(S, h)$ for a sample $S$ for some specific values of $h$
like other authors did (see Kawashima and Suzuki\cite{kawashima}). We can
evaluate the complete
piecewise constant function $m(S, h)$ for each sample. We do this by
starting at $h=0$, computing the ground state, and finding the next
increased value of $h$ for which the current ground state loses
optimality using a sensitivity analysis technique\cite{sensitivity}.
At that point we compute the new ground state. We do this up to a
given field strength or until saturation occurs.

This technique gives us the (averaged) function $m_L(h)$ for each
system size $L$ with any arbitrary resolution. 
We used systems of sizes $L \in \{10, 20, 30, 40, 50, 60\}$ and
computed the ground states for $\lceil{\frac{10^5}{L^2}}\rceil$ samples
for each size $L$.
We judged the data collapse in a plot $m_L L$ versus $L h^{1/\delta}$
as shown in Fig.\ 5 by visual inspection and using
cubic spline interpolation. In the figure we have included some
error bars for the $L=50$ and the $L=60$ curves to show the typical
errors. The errors decrease with decreasing system size, because of the
increasing number of samples.

We obtained the best data collapse at 
\begin{equation}
1/\delta=0.675\pm0.005\quad\Rightarrow\quad\delta=1.481\pm 0.011
\end{equation}
a value that agrees well with the result of the ground state
magnetization study by Kawashima and Suzuki \cite{kawashima}.  This
value together with estimate $y=0.281$ (\ref{yest}) from the defect
energy calculation implies that the scaling hypothesis (\ref{dypred}) is
significantly violated. Since we estimated $\nu$ directly in a Monte
Carlo simulation we conclude that it also not legitimate to infer from
the magnetic exponent $\delta$ via (\ref{dypred}) and (\ref{dypred})
that the thermal exponent $\nu$ should be close to $2$ as found in
\cite{liang,kawashima2}.

\section{Chaos exponent}

\label{Chaos_sec}

One of the peculiar features of spin glasses is their extreme
sensitivity with respect to parameter changes \cite{bm_chaos},
like small temperature, field or coupling variations.
For the ground state properties this means that a slight perturbation
of the initial set of couplings leads to a complete reorganization of
the original ground state over a length scale that depends on the
strength of the perturbation. This overlap length is expected to
behave as follows \cite{bm_chaos,imry}:

Let us modify the interactions by replacing each coupling $J_{ij}$ by
$J_{ij}'=J_{ij}+\delta K_{ij}$. Here $K_{ij}$ is again a Gaussian 
distributed random number with variance one and the parameter $\delta$
measures the strength of the perturbation. The comparison
of the energy balance $\Delta E_{\rm defect}$ for turning over a 
connected spin cluster of linear extent $L$ with the change of the ground 
state energy $\Delta E_{\rm random}$ induced by the random variation
of the couplings yields an estimate for the length scale beyond which
the original ground state is unstable with respect to the
perturbation. $\Delta E_{\rm defect}$ is simply the defect energy, which
is proportional to $JL^y$ (see section \ref{defect_sec}). The 
contribution to $\Delta E_{\rm random}$ coming from the $L^{d_S}$ interface 
spins of the cluster ($d_S$ being the fractal dimension of the interface) 
is proportional to $\delta L^{d_S/2}$. Thus for $L>L^*(\delta)$ with
\begin{equation}
L^*(\delta)\sim(J/\delta)^{-1/\zeta}\qquad{\rm with}\quad\zeta=d_S/2-y
\end{equation}
we have $\Delta E_{\rm random}(L)>\Delta E_{\rm defect}(L)$ and flipping
of clusters is favored by the perturbation. Thus the ground state
configurations in the original (denoted by $S_i$) and the perturbed
sample (denoted by $S_i'(\delta)$) become uncorrelated for distances
larger than $L^*$.

This statement can be quantified by studying the overlap correlation
function
\begin{equation}
C_\delta(r)=\left[\frac{1}{N}\sum_{i=1}^N\;
S_i\,S_{i+r}\;S_i'(\delta)\,S_{i+r}'(\delta)\;\right]_{\rm av}.
\end{equation}
According to the above mentioned argument one expects in the limit 
$N\to\infty$ a scaling form
\begin{equation}
C_\delta(r)\sim\tilde{c}(r\delta^{1/\zeta}).
\label{chaos_scale}
\end{equation}

\begin{figure}[hbt]
\epsfxsize=\columnwidth\epsfbox{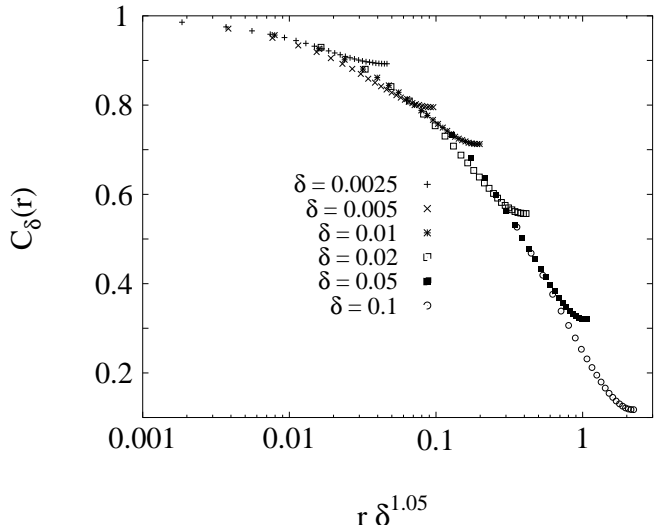}
\caption{Scaling plot of the overlap correlation function
  $C_\delta(r)$ versus $r/L^*$ with $L^*=\delta^{-1/\zeta}$.
  The best data collapse (for data confined to $r<L/4$) is obtained for
  $1/\zeta=1.05$. The system size is $L=50$ and the data are averaged
  over 400 samples. These were obtained by creating $>80$ reference 
  instances and creating 5 random perturbations of strength $\delta$ for each.}
\label{overlap}
\end{figure}

In figure \ref{overlap} we show the result of our calculation of the
overlap correlation function $C_\delta(r)$. We fixed the system size
to $L=50$, for which reason one has to neglect the data points for
$r>L/4$ (note the upwards bending due to the periodic boundary
conditions). For the rest of the data we obtain the best data collapse
for
\begin{equation}
1/\zeta=1.05\pm0.05\qquad{\rm i.e.}\quad\zeta=0.95\pm0.05
\end{equation}
which agrees well with the estimate from Bray and Moore
\cite{bm_chaos} obtained in a different way and by considering smaller
system sizes. With the value for $y$ we reported in section
\ref{defect_sec} the fractal dimension of the interface of an
excitation is given by $d_S=1.34\pm0.10$.

In passing we mention that the dependency of $C(r)$ on distance $r$
is neither exponential nor algebraic: it can nicely be fitted with a
stretched exponential
\begin{equation}
C(r)\approx\exp(-r^a/b)+\exp(-(L-r)^a/b)
\label{fit}
\end{equation}
with fit parameters $a$ and $b$. For instance $\delta=0.1$ for $L=50$
yields $a=0.8$ and $b=0.75$. Since $a$ and $b$ seem to depend on the
perturbation strength $\delta$ we do not expect the form (\ref{fit})
to be universal.

\begin{figure}[hbt]
\epsfxsize=\columnwidth\epsfbox{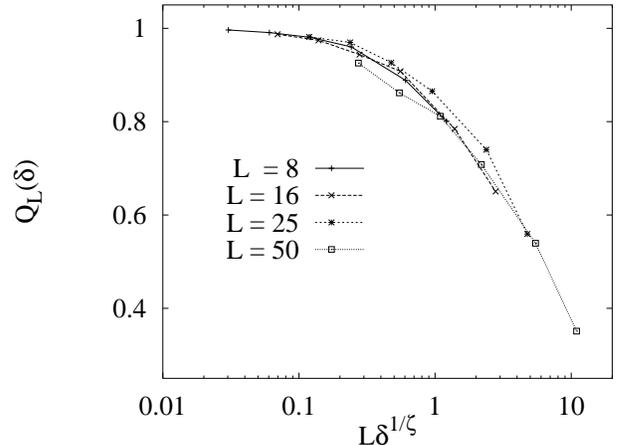}
\caption{Scaling plot of the ground state overlap $Q_L(\delta)$. The
  best data collapse is obtained with $1/\zeta=1.2$.}
\label{q_overlap}
\end{figure}

Defining $\xi_L(\delta)=\sum_{r=0}^L C_\delta(r)$ one expects from
(\ref{chaos_scale})
\begin{equation}
\xi_L(\delta)\sim\tilde{\xi}(L\delta^{1/\zeta})
\end{equation}
and in a more direct way for the ground state overlap \cite{bm_chaos}
$Q_L(\delta)=|\sum_{i=1}^{N} S_i\,S_i'(\delta)|$
\begin{equation}
Q_L(\delta)=\tilde{Q}(L\delta^{1/\zeta}).
\end{equation}
Note that $\xi_L(\delta)=Q_L^2(\delta)$. We show a finite size scaling
plot for $Q_L(\delta)$ in fig.\ {\ref{q_overlap}}, from which we
estimate $1/\zeta=1.2\pm0.1$. The quality of the data collapse is
good (cf.\ fig.\ 2 of ref.\ \cite{bm_chaos}).

\section{Summary}

With the help of an improved branch \& cut algorithm we were able to
reinvestigate the critical behavior of the two-dimensional Ising spin
glass model with a continuous bond distribution with much better
accuracy. We found that the stiffness exponent is given by
$y=-0.281\pm0.002$ implying a correlation length exponent of
$\nu=3.56\pm0.02$, which agrees well with our independent estimate
$\nu=3.6\pm0.2$ from Monte Carlo simulations. For the latter we
introduced a pseudo-ferromagnetic order parameter with the help of
exactly known ground states and analyzed its finite size scaling behavior
at non-zero temperatures.

We hope that our calculation settles the controversy regarding the
thermal exponent $\nu$ initiated by the cluster Monte Carlo study of
Liang \cite{liang} and the numerical transfer matrix study by
Kawashima et al.\ \cite{kawashima2}: Their values for $\nu$ are
substantially smaller than ours indicating a violation of the scaling
prediction by Bray and Moore \cite{bm_sc}. Our results for $y$ and
$\nu$ are clearly compatible with this scaling prediction $\nu=-1/y$.

Furthermore we determined exact ground states for systems within an
external field and from a finite size scaling analysis of the
magnetization we obtained an independent estimate for the magnetic
exponent $\delta=1.48\pm 0.01$. This confirms an earlier observation
\cite{kawashima} that there seems to be a disagreement between
the scaling theory \cite{bm_sc} predicting $\delta=1-y$ and the
numerical values obtained so far. In particular this discrepancy
does not fade away for larger system sizes, which we were able to
study here. Therefore our conclusion is that there must be a deeper
reason for this disagreement than some finite size effect which might
disappear if one only considers large enough system sizes.

Moreover, we calculated the overlap correlation function by perturbing
the bonds slightly in a random manner. We found a chaos exponent
$\zeta=0.95\pm0.05$ in agreement with earlier estimates from the analysis of
smaller system sizes.

Finally a few words concerning future perspectives: First we would
like to point out that in principle it is possible to improve the
system sizes {\it and} quality of statistics even further with the
algorithm we have at hand, provided we could simply run it on a
powerful parallel machine. However, our algorithm relies heavily on a
commercial linear problem solver for which we do not have a license to
run it on hundreds of processors of a e.g.\ Paragon XP/S10. On such a
machine we could possibly obtain an acceptable quality of statistics
for $L=100$, for which we can presently do only a few samples in
reasonable time on individual workstations.

As has been mentioned in the introduction, recently a finite
temperature phase transition in the site disordered Ising spin glass
has been reported \cite{remark}. Since the critical temperature is
pretty small, though, Monte Carlo studies might be hampered by
equilibration problems. Therefore this result could be put on a much
firmer base, if the stiffness exponent $y$ would indeed turn out to be
positive in this particular two-dimensional model and so signaling
the stability of the spin glass ordered phase for small, non-vanishing
temperatures. We intend to answer this question with our algorithm
soon.

Furthermore an obvious and highly rewarding step would be to perform
the same study in three dimensions. To calculate ground states for the
three-dimensional Ising spin glass model is an NP-complete problem and
the two-dimensional problem we have studied here is NP-complete, too
(note we have a continuous bond distribution, periodic boundary
conditions and an external field). However, although both questions
belong to the same class of {\it hard} combinatorial problems the
three-dimensional Ising spin glass is {\it much} harder, which means
that the operation count will be much higher: either the power of the
$L$, the system size, or the coefficient in the exponent will be
larger for three dimensions than for two dimension. Nevertheless we
are currently undertaking efforts in this direction, our progress in
this matter will be reported elsewhere.

\acknowledgements

We should like to thank A.\ Bray, C.\ De Simone, G. Reinelt, G.\ 
Rinaldi, D.\ Stauffer, and A.\ P.\ Young for various stimulating
discussions and valuable hints and remarks. The Monte Carlo
simulations have been done on the Paragon XP/S10 of the ZAM at KFA
J\"ulich.

\end{document}